# Mode coupling and resonance instabilities in quasi-two-dimensional dust clusters in complex plasmas

Ke Qiao,* Jie Kong, Jorge Carmona, Lorin S. Matthews and Truell W. Hyde†
Center for Astrophysics, Space Physics and Engineering Research, Baylor University, Waco, Texas 76798-7310, USA

Small quasi-two-dimensional (2D) dust clusters consisting of three to eleven particles are formed in an argon plasma under varying rf power. Their normal modes are investigated through their mode spectra obtained from tracking the particles' thermal motion. Detailed coupling patterns between their horizontal and vertical modes are detected for particle numbers up to seven and discrete instabilities are found for dust clusters with particle number $\geq 9$, as predicted in previous theory on ion-flow induced mode coupling in small clusters. The instabilities are proven to be induced by resonance between coupled horizontal and vertical normal modes.

Mode coupling is of fundamental importance across a variety of research fields including molecular chemical physics [1-2], glass transition physics [3], fiber optic physics [4], geophysics [5], and plasma physics [6]. Normal modes produced in an ideal linear system are independent of one another, i.e. no interaction or energy transfer occurs between modes. Although real-world systems are by nature nonlinear, many can be modeled as a linear system plus a nonlinear perturbation. This allows inspection of the interaction and/or energy transfer that exists between modes, which have been related to fundamental phenomena such as the intramolecular vibrational energy redistribution observed in polyatomic molecules [1-2] and increase in relaxation time of super-cooled liquids [3].

Complex (dusty) plasma physics has seen increased attention over the past two decades partly due to its ability to act as an excellent model for the study of condensed matter states at the kinetic level. Complex (dusty) plasma experiments typically examine dust particles levitated within the plasma sheath, where directional ion streaming occurs. This has allowed examination of mode couplings between longitudinal in-plane [7-8] and transverse out-of-plane dust lattice waves (DLWs) [9-11] in large two dimensional (2D) plasma crystals [12-19]. The ambient ion flow and resulting ion wakefield behind the particles [20-24] can dramatically enhance such coupling mechanisms from second (nonlinear) to first (linear) order [18], allowing detection even at low amplitude particle displacements. The subsequent increase in energy density is observed in the thermal fluctuation spectrum at the intersection between dispersion relations of the two DLWs, corresponding a resonance instability allowing melting of the plasma crystal at low neutral gas pressures. [16-19]

Dust clusters, on the other hand, are composed of a finite number of dust particles. Their unique characteristics allow them to exhibit ordered structures analogous to those observed for finite numbers of ions or electrons in traps [25,26]. Dynamic cluster properties are not described in terms of wave modes but in terms of normal modes with discrete frequencies. Coupling between horizontal (in plane) [26-29] and vertical (out-of-plane) normal modes [30] have been investigated theoretically for two to seven particles [31-35]. The coupling process has been shown to be enhanced by the ion wakefield but, unlike the continuous high energy density region at the intersection between the DLWs observed in large crystals, a complex coupling pattern exists between multiple horizontal and vertical modes even when the modes are not in resonance. Such coupling occurs only between modes having specified symmetries and obeying specific mode coupling rules, similar to the mode coupling seen in polyatomic molecules [1-2]. The instabilities created by the resonance between two coupled modes are discrete due to the fact that the normal modes are discrete in frequency [35].

In this paper, an experimental investigation of the normal modes occurring within complex plasma in 2D and quasi-2D dust clusters is reported. Examination of the mode spectra reveals detailed coupling patterns between vertical and horizontal modes and the predicted discrete resonance instabilities induced by these couplings [35] have been detected.

The experiment was carried out in a Gaseous Electronics Conference rf reference cell [36]. The cell contains two 8 cm-diameter electrodes separated by a distance of 1.9 cm. The lower electrode is powered at 13.56 MHz while the upper ring-shaped electrode and chamber are grounded. A 20 mm × 18 mm × 18 mm (height × length × width) glass box was placed on the lower electrode to create the confinement potential needed to establish the dust clusters. All experiments were conducted in argon plasma at 5 Pa employing rf powers between 0.8 W and 19 W. Melamine formaldehyde (MF) particles (mass density of 1.51 g/cm$^3$ and diameter of 8.89 ± 0.09 μm) were used.

Under these conditions, 2D or quasi-2D dust clusters consisting of three to eleven MF dust particles were formed within the glass box at the balance point between the gravitational force, interparticle force and electric field forces created by the lower electrode and the walls of the box. Particles were illuminated using a horizontally fanned laser sheet. Once the dust clusters were stabilized, top and side views were recorded employing CCD cameras running at 60 frames per second. The resulting image series were analyzed employing particle tracking to obtain each particle's position and velocity within each frame. For the small particle-number clusters used in this experiment, the particles seen in both top and side views were matched using simple geometry. Normal mode spectra were then obtained employing the technique described by Melzer [27], where a time series of particle velocities is projected onto the direction of the eigenvectors corresponding to each pure mode (i.e. the eigenmodes calculated assuming a Yukawa interaction). The normal mode power spectra were then obtained through Fourier transformation.

Fig. 1 (a, b) shows the normal mode spectra for three and four particle 2D clusters at rf powers of 2.32 W and 1.60 W, respectively. The $2N$ (with $N$ being the particle number) spectral lines from the left correspond to the horizontal modes and the $N$ lines from the right correspond to the vertical modes. (To our knowledge, vertical modes in clusters have never been previously observed experimentally.) These modes can be symbolized using order integers ($m$, $n$) following the conventions given in [30, 35], where the order integer $m = 0, 1, 2, ...$ are determined from the circumferential period of the oscillation patterns (eigenvectors) and $n = 1, 2, ...$ acts as the frequency index, Fig. 1(e, f). In agreement with previous research [37, 38], the vertical sloshing mode frequency change dramatically with change in rf power while the horizontal sloshing mode frequencies remain almost constant. Thus, as the rf power is decreased, the vertical modes' frequencies decrease while the horizontal modes remain constant in frequency, allowing the overlap between horizontal and vertical mode branches to be controlled and resonances between the normal modes to be investigated.

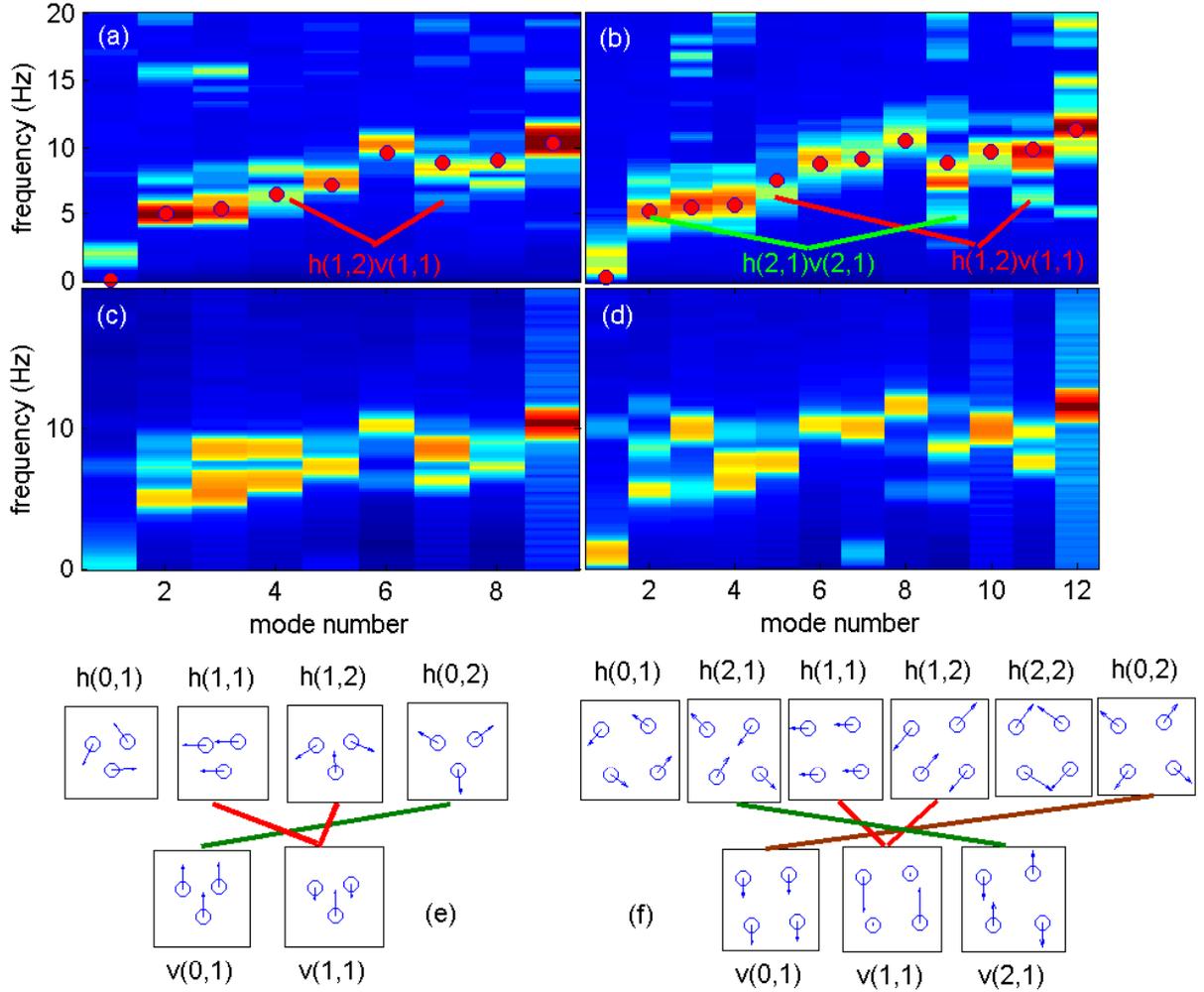

Fig. 1. (a-b) Normal mode spectra obtained experimentally for three and four particle clusters at rf powers of (a) 2.32 W and (b) 1.60 W. Dots represent theoretically calculated mode frequencies. (c-d) Normal mode spectra obtained from simulation with parameters matching those measured in (a) and (b). (e-f) Mode eigenvectors for clusters corresponding to (a) and (b), where the horizontal (vertical) modes are shown at top (bottom). The solid lines shown in (a-b) and (e-f) connect coupled modes.

Since the interparticle potential in the horizontal direction is known to be well described by a Debye-Hückel, or Yukawa potential [39, 40], the particle charge and screening length can be estimated by fitting the horizontal spectral lines with their corresponding mode frequencies (represented by the solid dots in Fig. 1(a,b)) as theoretically calculated employing the dynamic matrix for a parabolically confined Yukawa system [41, 42] (the confinement coefficients are determined by the sloshing mode frequencies), combined with the force balance equation. Using this technique, the charge on the particles for this experiment was found to be $q$ = 11600±1000 e with a screening length in the range $\lambda_D$ = 200~400 μm.

Coupling between horizontal and vertical modes can be identified following the method outlined in [34, 35] by detecting multiple modes having the same spectral frequency. For example in a three particle cluster, Fig. 1(a), at the frequency of the horizontal "kink" mode

($h$(1,2), mode number 4), f ≈ 6 Hz, a light spectral line corresponding to the vertical relative mode ($v$(1,1), mode number 7) can be seen. This implies that this mode is no longer a pure $h$(1,2) mode, but rather a mode with mixed polarization, consisting of both $h$(1,2) and $v$(1,1) components. Similarly, at the $v$(1,1) mode frequency f ≈ 8.5 Hz, a spectral line corresponding to $h$(1,2) mode can be found. Thus, the $h$(1,2) and $v$(1,1) modes are mutually coupled to each other. With this method, two mutual couplings, $h$(1,2)$v$(1,1) and $h$(2,1)$v$(2,1) for the four particle cluster were identified, as predicted in previous numerical simulations [35].

The primary difference between current experimental data and previous simulations is that the confinement in the $x$ and $y$ directions in the experiment are slightly different, creating a loss of degeneracy between some modes. For this reason an $N$-body code [35] was employed to simulate conditions having slightly different $x$ and $y$ confinements. Following [35], the ion wakefield is taken into account using the point charge model [12-15, 32-35, 37, 43]. Figures 1(c,d) show the normal mode spectra obtained from this simulation, with the particle charge, screening length, and $x$, $y$ and $z$ confinement equal to those measured under conditions corresponding to Figs.1(a,b). The magnitude and distance of the point charge below any actual particle are set to be $q/2$ and $\lambda_D/3$, respectively, values chosen to be in agreement with the majority of current theoretical ion wakefield models [20-24]. Comparing Figs. 1(a,b) with Figs. 1(c,d) it is seen that an excellent match is found between the coupling patterns obtained from experiment and simulation, even for this simple model of the ion wakefield.

Unlike three and four particle dust clusters, which form hollow triangular or square structures, clusters with particle number 5-8 exhibit a stable structure with one particle at the center surrounded by $N$-1 particles (for $N$ = 5, this is a metastable structure). Under the current experimental setup, the center particle is *always* positioned slightly below the outer particles with this becoming more pronounced as the rf power decreases. Therefore, for 5-8 particles, the rf power cannot be lowered enough to allow overlap between the horizontal and vertical spectral branches without loss of the cluster's 2D nature.

Nevertheless, detection of horizontal-vertical mode couplings is still feasible for 5-8 particle clusters while retaining their 2D characteristics at higher rf powers (> 7 W), where the horizontal and vertical branches are well separated in their mode spectra. As a representative example, Fig. 2(a) shows the mode spectra for a seven-particle cluster. Since the coupling strength between the horizontal and vertical modes is now much weaker, the overall intensity of the mode spectra has been enhanced in order to bring out the fainter details. In this case, spectral lines caused by mode couplings on the horizontal side (mode numbers 1-2$N$) cannot be detected since the horizontal thermal motion (with average particle velocity greater than $2 \times 10^4$ m/s) and external perturbation (such as an integral rotation of the cluster) is strong enough to dominate the coupling signals. On the other hand, there is less thermal motion (with average particle velocity less than $1.5 \times 10^4$ m/s) and external perturbation in the vertical direction. Thus clear spectral lines can be seen on the vertical side (mode numbers 2$N$+1 to 3$N$) allowing the identification of the mode couplings. The theoretically predicted $h$(1,2)$v$(1,1), $h$(2,1)$v$(2,1), $h$(3,1)$v$(3,1), $h$(1,3)$v$(1,1), $h$(2,2)$v$(2,1) and $h$(0,2)$v$(0,2) mode couplings can be seen clearly in Fig. 2(a) and an excellent match with the coupling pattern obtained from simulation, Fig. 2(b) is again found.

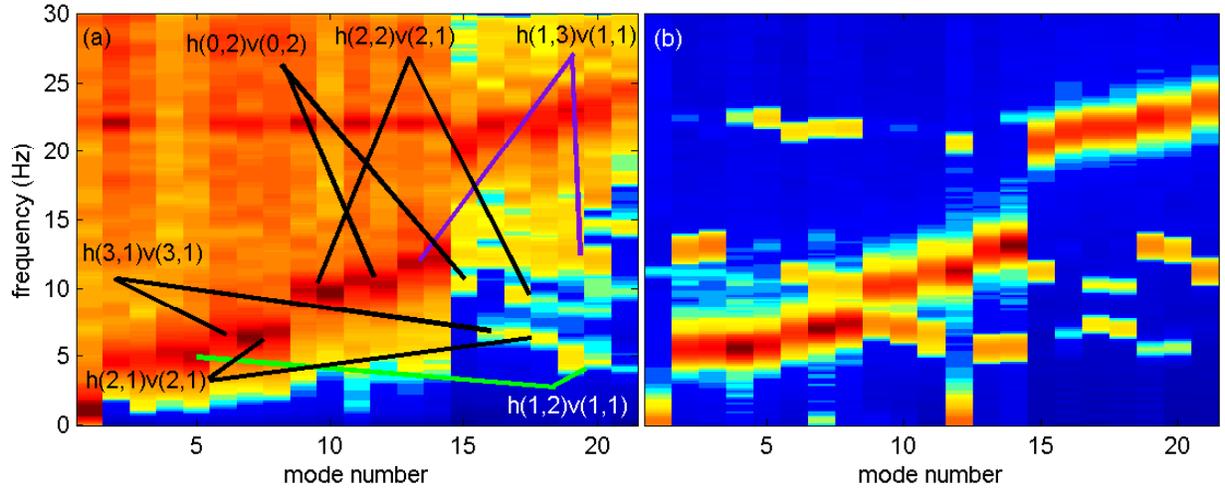

Fig. 2. (a) Normal mode spectra for a seven-particle cluster at rf power of 17.2 W. The solid lines shown connect coupled modes. (b) Normal mode spectra obtained from simulation with parameters matching those measured in (a).

As defined in [35], mode couplings can be classified as one of three types. Type I represents a non-mutual coupling with one of the coupled modes being a sloshing mode, while Type II and III both represent mutual couplings. Types I and III create resonance instabilities while Type II does not. Type I coupling was not detected in the current experiment, in agreement with the prediction that this type of coupling is much weaker than Types II and III (i.e., the spectral lines are easily dominated by noise).

Simulation in [35] also predicted no instability-inducing (Type III) coupling for clusters with $N < 5$, in agreement with our observation. However, the resonance instabilities predicted for clusters with $5 \leq N \leq 8$ were not observed in the current experiment. This can be explained by the large vertical spread for 5-8 particle clusters as mentioned above, making the theory employed in [35], which is based on a 2D model, no longer applicable.

On the other hand, instabilities for larger clusters with $N > 8$ *were* observed. One characteristic of the resonance instabilities predicted, which is unique to small clusters, is their discreteness [35]. This discreteness can clearly be seen in the current experiment as the rf power changes. For example, a nine particle dust cluster remains stable for rf powers > 15 W. However as the rf power is decreased to approximately 13 W, an instability occurs, as identified by particle trajectories, Fig. 3(d, e). As the power is decreased further, the system gradually returns to a stable configuration at approximately 9 W, reaching a second instability near 6 W. Finally, a third stable regime appears at a plasma power of approximately 4 W, followed by a third instability at approximately 2.7 W, which eventually splits the system into two layers losing its quasi-2D structure.

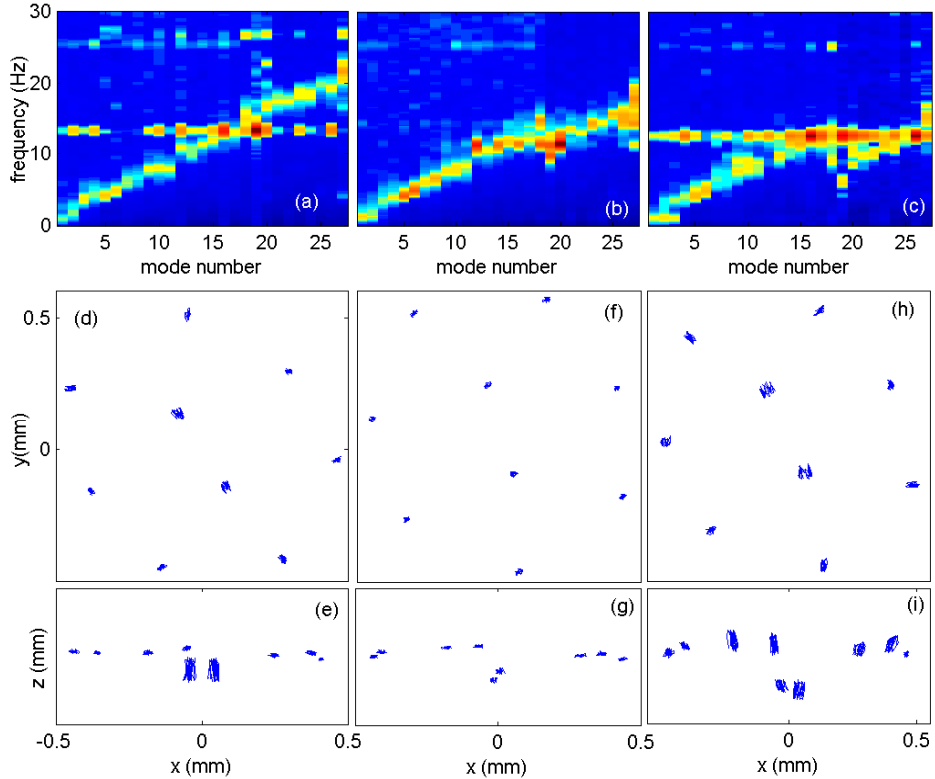

Fig. 3. (a-c) The mode spectra of a nine-particle cluster at rf powers of (a) 11.5 (b) 7.5 and (c) 5.2 W. (d-i) The particle trajectories from top and side views for clusters in (a-c).

Fig. 3 shows the mode spectra for a representative nine-particle cluster at the onset of the first two of these instabilities, the stable state between them and the particle trajectories viewed from the top and side. The instabilities, which can be seen in the particle trajectories, are reflected in the increase in both the horizontal and vertical motion. This is a signature of instabilities created by resonance between two mutually coupled horizontal and vertical modes. This simultaneous excitation of horizontal and vertical oscillations also agrees with both previous theoretical results and experimental observations for large plasma crystals [15-19].

Comparing mode spectra for stable and unstable clusters, it can be seen that the latter exhibit a clear cascade of energy density passing through the horizontal and vertical branches, while the former shows no such cascades. Meanwhile, mode spectra for the unstable clusters show "hot spots" at the intersections between the cascade and the horizontal and vertical branches, exhibiting dramatic energy density enhancements at a horizontal and a vertical mode with the same frequency (e.g., mode number 16 and 19 in Fig. 3(a) and mode number 16 and 26 in Fig. 3(c)), direct evidence that these instabilities are caused by a resonance between two coupled modes. A closer look at the excited modes appearing in Fig. 3(a) is provided by comparing their eigenvectors (oscillation patterns) with the particle trajectories (Fig. 4). Both the horizontal and vertical trajectories of the particles match their directions in the eigenvectors, showing that the particle motion is indeed dominated by the excited modes. The fact that cascades appear in the mode spectra for clusters with as few as nine particles indicates that this is not a phenomenon unique to large systems [17-19]. It also appears to agree with the argument that such cascades are a nonlinear effect given the second harmonic seen at double the resonant frequency. Thus, the cascade could be caused by a chain resonance between multiple horizontal and vertical coupled modes.

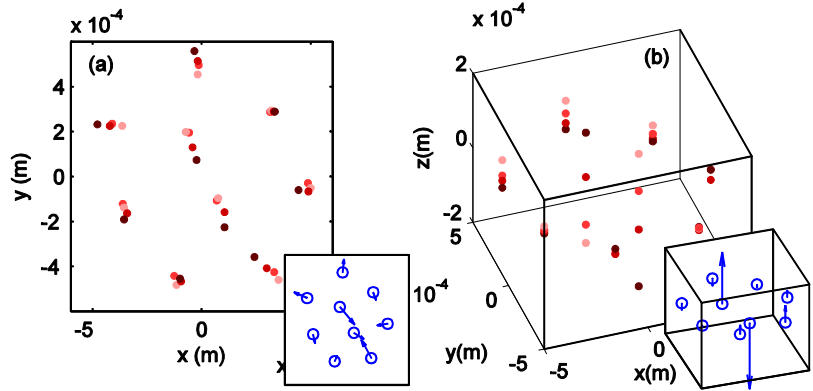

Fig. 4. The (a) horizontal and (b) vertical particle trajectories for a nine particle cluster at rf power of 11.5 W. Displacements from equilibrium positions are magnified by a factor of five in order for the trajectories to be seen more clearly. Particle positions as a function of time are indicated by the intensity of the markers, going from light to dark. Insets in (a) and (b) are the eigenvectors of the excited horizontal and vertical modes, respectively.

The vertical spread within the clusters, which is a measure of deviation from an ideal 2D system, creates the primary experimental error. As mentioned previously, the vertical spread found in 5-8 particle clusters may well explain the fact that no instabilities were observed for clusters of this size. For larger clusters (e.g. nine particles), the vertical spread can increase to as much as 15% of the horizontal diameter at the lowest rf power. This deviation creates effects not easily explained by a 2D mode coupling model. For example, the central particles, which are positioned lower than the plane defined by the particles comprising the rest of the cluster, vibrate more strongly than do the outer particles, Fig. 3(e,i) due to the so-called Schweigert mechanism [43]. This results in an abnormal energy density enhancement ("hot spot") such as the 19th mode in Fig. 3(c) and which is not at the intersection between the cascade and vertical mode branch.

In summary, 2D or quasi-2D dust clusters consisting of three to eleven particles formed in an argon plasma across a rf power regime between 0.8 and 19 W were examined with individual particle motions tracked to obtain the clusters' normal mode spectra. Coupling patterns between horizontal and vertical modes were detected and found to be in excellent agreement with theoretical predictions for clusters having particle numbers up to seven. Discrete instabilities, also predicted theoretically, were observed for dust clusters with particle numbers $N \geq 9$ as the rf power varied. Inspection of the resulting normal mode spectra and comparison between particle trajectories with mode eigenvectors both indicate these instabilities are caused by resonances between coupled horizontal and vertical modes. This observation of mode coupling patterns and resulting discrete instabilities not only serves as direct proof of the theory for ion-flow induced mode coupling in small clusters, but also opens the possibility for using dust clusters to investigate mode interactions on the kinetic level in other finite systems, for example, in the semi-classical study of polyatomic molecules.

We thank Peter Hartmann for helpful discussions.

*ke_qiao@baylor.edu
†truell_hyde@baylor.edu

[1] T. Uzer, Phys. Reports, 199, 73 (1991).
[2] B. C. Dian, A. Longarte, and T. S. Zwier, Science 296, 2369 (2002).
[3] K. N. Pham *et al*. Science 296, 104 (2002).
[4] Nenad Bozinovic *et al*. Science 340, 1545 (2013).
[5] J. Resovsky and M. Ritzwoller, Geophys. Res. Lett. 22, 2301 (1995).
[6] İ. Ü. Uzun-Kaymak *et al*, Europhysics Lett. 85, 15001 (2009).
[7] A. Homann *et al*., Phys. Lett. A 242, 173 (1998).
[8] S. Nunomura, J. Goree, S. Hu, X. Wang, A. Bhattacharjee, and K. Avinash, Phys. Rev. Lett. 89, 035001 (2002).
[9] S. V. Vladimirov, Physica A 315, 222 (2002).
[10] K. Qiao and T.W. Hyde, Phys. Rev. E 68, 046403 (2003).
[11] K. Qiao and T.W. Hyde, Phys. Rev. E 71, 026406 (2005).
[12] A. V. Ivlev and G. Morfill, Phys. Rev. E, 63, 016409 (2000).
[13] S. K. Zhdanov, A. V. Ivlev and G. E. Morfill, Physics of Plasmas 16 083706 (2009).
[14] T. B. Rocker, A. V. Ivlev, R. Kompaneets and G. E. Morfill, Physics of Plasmas 19, 033708 (2012).
[15] T. B. Röcker, A. V. Ivlev, S. K. Zhdanov, and G. E. Morfill, Phys. Rev. E 89, 013104 (2014).
[16] A. V. Ivlev, U. Konopka, G. Morfill and G. Joyce, Phys. Rev. E 68, 026405 (2003).
[17] L. Couëdel, V. Nosenko, A. V. Ivlev, S. K. Zhdanov, H. M. Thomas and G. E. Morfill, Phys. Rev. Lett. 104, 195001 (2010).
[18] B. Liu, J. Goree and Y. Feng, Phys. Rev. Lett. 105, 085004 (2010).
[19] L. Couëdel, S. K. Zhdanov, A. V. Ivlev, V. Nosenko, H. M. Thomas and G. E. Morfill, Phys. Plasmas, 18, 083707 (2011).
[20] S. V. Vladimirov and M. Nambu, Phys. Rev. E 52 R2172 (1995).
[21] F. Melandsø and J. Goree, Phys. Rev.E 52 5312 (1995).
[22] O. Ishihara and S. V. Vladimirov, Phys. Plasmas 4, 69 (1997).
[23] M. Lampe, G. Joyce, G. Ganguli and V. Gavrishchaka, Phys. Plasmas 7, 3851 (2000).
[24] R. Kompaneets, U. Konopka, A. V. Ivlev, V. Tsytovich and G. Morfill, Phys. Plasmas 14 052108 (2007).
[25] V.M. Bedanov and F. Peeters, Phys. Rev. B 49, 2667 (1994).
[26] V.A. Schweigert and F. Peeters, Phys. Rev. B 51, 7700 (1995).
[27] A. Melzer, Phys. Rev. E 67, 016411 (2003).
[28] Minghui Kong, B. Partoens, A. Matulis and F. M. Peeters, Phys. Rev. E 69, 036412 (2004).
[29] T. E. Sheridan and W. L. Theisen, Phys. Plasmas 13, 062110 (2006).
[30] Ke Qiao and Truell W. Hyde, IEEE transactions on plasma science 36, 2753 (2008).
[31] R. Kompaneets, S. V. Vladimirov, A. V. Ivlev, V. Tsytovich and G. Morfill, Phys. Plasmas 13, 072104 (2006).
[32] V. V. Yaroshenko, S. V. Vladimirov and G. E. Morfill, New Journal of Physics 8, 201 (2006).
[33] K. Qiao, L. S. Matthews and T. W. Hyde, IEEE transactions on plasma science 38, 826 (2010).

[34] Ke Qiao, Jie Kong, Zhuanhao Zhang, Lorin S. Matthews and Truell W. Hyde, IEEE transactions on plasma science, 41 745 (2013).
[35] Ke Qiao, Jie Kong, Eric Van Oeveren, Lorin S. Matthews, and Truell W. Hyde, Phys. Rev. E 88, 043103 (2013)
[36] V. Land, B. Smith, L. Matthews, and T. W. Hyde, IEEE Trans. Plasma Sci. 38, 4 (2010).
[37] V. Steinberg, R. Sütterlin, A. V. Ivlev and G. Morfill, Phys. Rev. Lett. 86, 4540 (2001).
[38] A. A. Samarian, S. V. Vladimirov, and B.W. James, Phys. Of plasmas 12, 022103 (2005).
[39] U. Konopka, G. E. Morfill and L. Ratke, Phys. Rev. Lett. 84, 891 (1999).
[40] Z. Zhang, K. Qiao, J. Kong, L. Matthews and T. Hyde, Phys. Rev. E 82, 036401 (2010).
[41] B. Liu, K. Avinash and J. Goree, Phys. Rev. E 69, 036410 (2004).
[42] E. B. Tomme, B. M. Annaratone and J. E. Allen, Plasma Sources Sci. Technol. 9, 87 (2000).
[43] V. A. Schweigert et al., Phys. Rev. E 54, 4155 (1996).